\documentclass[prl,twocolumn,epsfig,epsf]{revtex4-2}
\usepackage{
amsmath,amsthm,amsfonts,amscd,mathrsfs,pifont,bm}
\usepackage{eucal,graphicx,color,hyperref}
\usepackage{amssymb,float,ulem}
\usepackage{tikz}
\usepgflibrary{arrows.meta}

\usepackage{booktabs}
\usepackage{multirow}

\usepackage{pgfplots}
\pgfplotsset{height=6cm,width= \columnwidth,compat=1.17}

\newcommand{\bftab}{\fontseries{b}\selectfont}

\newcommand{\adddot}[4]{
  \draw [fill,red] (-#1:1) circle [radius=#4] node [#3] {#2};
}
\newcommand{\addtic}[4]{
	\draw [very thick] (-#1:1cm-0.5*#4) -- (-#1:1cm) node [#3] {#2} --  (-#1:1cm+0.5*#4);
	\def\lastangle{#1}
}

\newcommand{\addsegcircle}[7]{
    \draw [very thick,rotate=-\lastangle] (0:1) arc [start angle=0, end angle=-#1+\lastangle, radius=1];	
	\draw [dashed,thick,rotate=-\lastangle,<->] (0:#7) arc [start angle=0, end angle=-#1+\lastangle, radius=#7];
	\draw (-0.5*\lastangle-0.5*#1:1cm)  (-0.5*\lastangle-0.5*#1:#7-0.05cm) node [#5] {#4};
	\addtic{#1}{#2}{#3}{#6}
}

\newcommand{\adddotted}[5]{
    \draw [very thick,rotate=-\lastangle] (0:1) arc [start angle=0, end angle=-#1+\lastangle, radius=1];	
    \draw [dotted,rotate=-\lastangle] (0:#5) arc [start angle=0, end angle=-#1+\lastangle, radius=#5];	
	\draw [very thick] (-#1:1cm-0.5*#4) -- (-#1:1cm) node [#3] {#2} --  (-#1:1cm+0.5*#4);
	\def\lastangle{#1}
}

\usepackage{esint}

\makeatletter
\@for\next:={int,iint,iiint,iiiint,dotsint,oint,oiint,sqint,sqiint,
  ointctrclockwise,ointclockwise,varointclockwise,varointctrclockwise,
  fint,varoiint,landupint,landdownint}\do{%
    \expandafter\edef\csname\next\endcsname{%
      \noexpand\DOTSI
      \expandafter\noexpand\csname\next op\endcsname
      \noexpand\ilimits@
    }%
  }
\makeatother

\newcommand{\beginsupplement}{%
        \setcounter{table}{0}
        \renewcommand{\thetable}{S\arabic{table}}%
        \setcounter{figure}{0}
        \renewcommand{\thefigure}{S\arabic{figure}}%
     }

\newcounter{supplsection}
\newcommand{\supplsection}[1]{\refstepcounter{supplsection}\section{\thesupplsection. #1}}

\begin{document}

\title  {Statistics of local level spacings in single- and many-body quantum chaos}

\author{Peng Tian${}^{1}$}
\altaffiliation[Present address: ]{Laboratoire Jean Alexandre Dieudonn\'e, Universit\'e C\^ote d'Azur, 06108 Nice, France.}
\author{Roman Riser${}^{1,2}$}
\altaffiliation[Present address: ]{Department of Computer Science, Texas Tech University, Lubbock, TX 79409, USA.}
\author{Eugene Kanzieper${}^{1,3}$}
\affiliation{${}^{1}$School of Mathematical Sciences, Holon Institute of Technology, Holon 5810201, Israel}
\affiliation{${}^{2}$Department of Physics and Research Center for Theoretical Physics and Astrophysics, University of Haifa, Haifa 3498838, Israel}
\affiliation{${}^{3}$Department of Physics of Complex Systems, Weizmann Institute of Science, Rehovot 7610001, Israel
       \\
       \\
       \rm{(Received 13 August 2023; revised 16 November 2023; accepted 22 April 2024; published 29 May 2024)}
}

\begin  {abstract}
We introduce a notion of {\it local level spacings} and study their statistics within a random-matrix-theory approach. In the limit of infinite-dimensional random matrices, we determine {\it universal sequences} of mean local spacings and of their ratios which uniquely identify the global symmetries of a quantum system and its internal -- chaotic or regular -- dynamics. These findings, which offer a new framework to monitor single- and many-body quantum systems, are corroborated by numerical experiments performed for zeros of the Riemann zeta function, spectra of irrational rectangular billiards and many-body spectra of the Sachdev-Ye-Kitaev (SYK) Hamiltonians.\\
\newline
Published in: \href{https://doi.org/10.1103/PhysRevLett.132.220401}{\textcolor{blue}{Phys. Rev. Lett. {\bf 132}, 220401 (2024)}}
\end{abstract}

\maketitle

\textit{\textbf{Introduction.}}---There is a broad consensus, based on a vast amount of experimental, numerical and theoretical evidence~\cite{GMGW-1998,S-1999,H-2001}, that a universal statistical behavior of single-particle quantum systems correlates with the nature -- chaotic or regular -- of their underlying classical dynamics. In this context, two exemplary universality classes have been identified in quantum chaology.

The Wigner-Dyson universality class accommodates generic quantum systems which are fully chaotic in the classical limit. According to the Bohigas-Giannoni-Schmit (BGS) conjecture \cite{BGS-1984,MK-1979,CVG-1980,B-1981,Z-1981}, spectral fluctuations of highly excited energy levels in such systems -- exhibiting long-range correlations
and local repulsion -- are governed by global symmetries rather than by system peculiarities and are accurately described by the infinite-dimensional random matrix theory~\cite{M-2004,PF-book}. On the contrary, generic quantum systems whose classical dynamics is integrable belong to a different -- Poisson -- universality class as was first conjectured by Berry and Tabor \cite{BT-1977}. Spectral fluctuations therein are radically different from those in Wigner-Dyson spectra, with energy levels being completely uncorrelated.

Ever since the invention of the random matrix theory, a variety of statistical indicators have been devised to study fluctuations in spectra of bounded quantum systems. They include the number variance~\cite{M-2004,Rem-NV}, distribution of spacings between consecutive~\cite{JMMS-1980,TW-1993} and nearest-neighbor~\cite{FO-1996} eigenlevels and, more recently, the power spectra of eigenlevels~\cite{RGMRF-2002,ROK-2017,ROK-2020,RK-2023} and spacings~\cite{O-1987,RTK-2023}. Apart from the number variance, the aforementioned statistical measures come with a caveat: they cannot be applied directly to the raw spectra. To detect the spectral universality, an influence of system-specific mean level density has to be eliminated first from measured sequences of energy levels by means of the unfolding procedure~\cite{M-2004}.

On the other hand, rapidly developing studies of quantum chaos in interacting many-body systems~\cite{OH-2007,ABGR-2013,GGV-2016,AWGBG-2017,BKP-2018,LVG-2020,SE-2020,MMTA-2021,RUT-2022,SLHC-2023,ACMORS-2023} (with or without the classical limit) have generated demand in alternative statistical tests which do not require a knowledge of local density of states, make the spectral unfolding redundant and thus allow for a more transparent and accurate comparison with experiments. These criteria are met by the $r$-statistics which deals with the {\it ratio}~\cite{OH-2007,ABGR-2013,Rem-ratio} of two consecutive level spacings~\cite{Rem-CP,SRP-2020}. 
First proposed in the numerical study by Oganesyan and Huse~\cite{OH-2007} and later handled analytically by Bogomolny and collaborators~\cite{ABGR-2013}, the $r$-statistics has two important advantages over traditional spectral fluctuation measures: the ratio of consecutive spacings not only is independent of the local density of states but also incorporates a non-trivial information about their correlations~\cite{Rem-corr,BLS-2001,RTK-2023}.

In this Letter, we introduce a notion of {\it local level spacings (LLS)}, study their fluctuational properties, and argue that several statistical measures associated with local spacings are particularly useful for {\it data analysis of raw spectra}. (We pinpoint the reader to our first and second main results and to the universal number sequences highlighted in Tables~\ref{local-means-table} and \ref{table-SYK-ratios}.) Apart from offering a meaningful alternative to the $r$-statistics and equipping the field with an independent tool for monitoring spectra of many-body quantum systems, the new statistics is of interest in its own right: intriguing and seemingly counter-intuitive properties of local level spacings can naturally be interpreted in terms of the famous `inspection paradox'~\cite{MH-2020,F-1971,PKR-2022} in probability theory. Our findings are corroborated by extensive numerical experiments performed for spectra of large-dimensional random matrices (Test~I), eigenlevels of rectangular billiards and nontrivial zeros of the Riemann zeta function (Test~II), and many-particle spectra of the SYK Hamiltonians (Test~III).

\textit{\textbf{Dyson's circular triad.}}---To set the stage, 
 we turn to a random-matrix-theory setup provided by the family of Dyson's circular ensembles ${\rm C}\beta{\rm E}(N)$ defined by the joint probability density function (JPDF)~\cite{M-2004}
\begin{eqnarray}\label{CbetaE-m-JPDF}
 \!\!\!\! P_N^{{\rm C}\beta {\rm E}}({\bm \theta}) = \frac{\Gamma(1+\beta/2)^N}{\Gamma(1+\beta N/2)}
    \prod_{1\le j<k\le N} \!\! | e^{i\theta_j} - e^{i\theta_k} |^\beta,
\end{eqnarray}
where $\beta=1, 2$ and $4$ is the Dyson symmetry index. Translationally invariant JPDF Eq.~(\ref{CbetaE-m-JPDF}) describes a set of $N$ repulsively interacting points (eigen-angles) ${{\bm\theta}=\{\theta_1,\dots,\theta_N\}\in [0,2\pi)^N}$ confined to the unit circle. Being primarily of mathematical interest at finite $N$, an  infinite-dimensional version of ${\rm C}\beta{\rm E}(N)$ is of direct physical relevance. Indeed, as $N\rightarrow\infty$, its eigen-angles, measured in units of the mean level spacing $\Delta_N=2\pi/N$, describe~\cite{GMGW-1998} the bulk spectral fluctuations in `maximally chaotic' quantum systems.

\textit{\textbf{Spacings between consecutive eigenlevels.}}---In the context of the ${\rm C}\beta{\rm E}(N)$ model, traditional level spacing refers to the length $s_X$ of an arc between a pair $(\theta_X, \theta_{X +1})$ of {\it consecutive} eigen-angles chosen at {\it random} out of the ordered set ${\{0 \le \theta_1 \le \dots\le \theta_N < 2\pi\}}$. Here $X$ is a uniformly distributed discrete random variable taking the values $\{1,2,\dots,N\}$ and $\theta_{N+1}=\theta_1+2\pi$. Obviously, the mean level spacing equals $\Delta_N = 2\pi/N$. The distribution functions of consecutive level spacings are well studied in both the ${\rm C}\beta{\rm E}(N)$ setting~\cite{PF-book}
\begin{eqnarray}\label{PN-consecutive}
    p_N^{{\rm C}\beta{\rm E}}(s) = {\mathbb E}_{\bm \theta} {\mathbb E}_X \left[ \delta(s-s_X) \right],
\end{eqnarray}
and in the ${N\rightarrow\infty}$ scaling limit that produces three universal spacing distributions~\cite{FW-2000,F-2005}
\begin{eqnarray} \label{N-infinity-lim}
    p^{(\beta)}(s) = \lim_{N\rightarrow\infty} \frac{2\pi}{N} p_N^{{\rm C}\beta{\rm E}}\left(\frac{2\pi}{N}s\right)
\end{eqnarray}
which are of central interest in numerous physical applications. Here, ${\mathbb E}_{\bm \theta}$ and ${\mathbb E}_X$ denote averaging with respect to the random ${\rm C}\beta{\rm E}(N)$ spectrum and the random variable $X$, respectively.

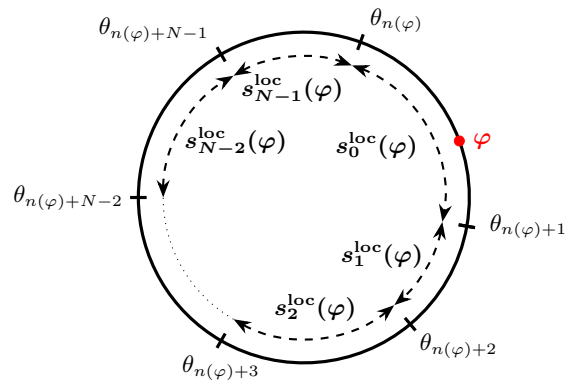
\begin{figure}[t]
\centering	
\begin{tikzpicture}[scale=2.2,>=Stealth] 
		\def\ltic{0.1cm} 
		\def\rarc{0.85cm} 
		\def\lastangle{0}
		\addtic{10}{$\theta_{n(\varphi)+1}$}{right=5pt}{\ltic}
		\addsegcircle{50}{$\theta_{n(\varphi)+2}$}{below right}{ $\bm{s_1^{\rm loc}(\varphi)}$ }{above left=-5pt}{\ltic}{\rarc}
		\addsegcircle{120}{$\theta_{n(\varphi)+3}$}{below=2pt}{ $\bm{s_2^{\rm loc}(\varphi)}$ }{above =0pt}{\ltic}{\rarc}
		\adddotted{180}{$\theta_{n(\varphi)+N-2}$}{left=2pt}{\ltic}{\rarc}
		\addsegcircle{240}{$\;\theta_{n(\varphi)+N-1}$}{above left =1pt}{ $\bm{s_{N-2}^{\rm loc}(\varphi)}$ }{below right=-5pt}{\ltic}{\rarc}
		\addsegcircle{290}{$\theta_{n(\varphi)}$}{above right=0pt}{ $\bm{s_{N-1}^{\rm loc}(\varphi)}$ }{below =0pt}{\ltic}{\rarc}
		\addsegcircle{370}{}{above right=-2pt}{ $\bm{s_{0}^{\rm loc}(\varphi)}$ }{below left =-3pt}{\ltic}{\rarc}
		\adddot{-20}{$\bm{\varphi}$}{right=2pt}{0.3mm}				
\end{tikzpicture}
\caption{Illustration of the definition of a set of local spacings $\{ s_0^{\rm loc}(\varphi;N), s_1^{\rm loc}(\varphi;N),\dots,s_{N-1}^{\rm loc}(\varphi;N)\}$ with respect to the fixed point $\varphi$ as introduced in the main text. The second argument ($N$) in the notation of local spacings was omitted for a better visual appearance.}\label{Fig_Local_Spacings}
\end{figure}

\textit{\textbf{Local level spacings (LLS).}}---Let us change the rules of the game, see Fig.~\ref{Fig_Local_Spacings} for an illustration. Instead of picking up a pair of consecutive eigen-angles at random, we fix a deterministic point ${\varphi \in [0, 2\pi)}$ on the circle {\it without a prior knowledge} of random positions of ${\rm C}\beta{\rm E}(N)$ eigen-angles, and seek for a pair ${(\theta_{n(\varphi)}, \theta_{n(\varphi)+1})}$ out of the ordered set of eigen-angles
${\{0 \le \theta_1 \le \dots\le \theta_N < 2\pi\}}$, where ${\theta_{N+1}=\theta_{1}+ 2\pi}$, such that $\theta_{n(\varphi)}<\varphi< \theta_{n(\varphi)+1}$. By construction, the arc, connecting $\theta_{n(\varphi)}$ and $\theta_{n(\varphi)+1}$, contains the fixed point $\varphi$. The arc length, to be denoted $s_0^{\rm loc}(\varphi;N)$, will be called {\it zeroth local spacing}. We then keep moving clockwise along the circle to identify the eigen-angles $\theta_{n(\varphi)+2},\dots, \theta_{n(\varphi)+N-1}, \theta_{n(\varphi)+N}$, where ${\theta_{n(\varphi)+N}=\theta_{n(\varphi)}+ 2\pi}$. Generically, the length of the $\ell$-th arc, connecting random points $\theta_{n(\varphi)+\ell}$ and $\theta_{n(\varphi)+\ell+1}$, to be denoted $s_\ell^{\rm loc}(\varphi;N)$, will be called the {\it $\ell$-th local spacing}~\cite{Rem-moments} for all $\ell=0,\dots,N-1$. Notice, that due to translational invariance of the JPDF Eq.~(\ref{CbetaE-m-JPDF}), the reference point $\varphi$ can be set to zero, or chosen at random, without loss of generality.

\textit{\textbf{Mean LLS.}}---A seemingly trivial question we would like to ask is this: What is the mean value $\langle s_\ell^{\rm loc}(\varphi;N) \rangle$ of the $\ell$-th LLS? The answer, which may sound counter-intuitive, is that $\langle s_\ell^{\rm loc} (\varphi;N) \rangle$ {\it differs} from the traditional mean level spacing ${\Delta_N=2\pi/N}$. We claim that~\cite{Rem-phi-independent}
\begin{subequations}\label{1st-finite-N}
\begin{eqnarray}\label{local-mean-N}
    \!\! \langle s_\ell^{\rm loc} (\varphi;N) \rangle = (1+\delta_{\ell,0}) \int_{0}^{2\pi} d\vartheta\, E_N^{{\rm C}\beta{\rm E}}(\ell;\vartheta),
\end{eqnarray}
where $E_N^{{\rm C}\beta{\rm E}}(\ell;\vartheta)$ is the probability to observe exactly $\ell$ eigen-angles in an arc of length $\vartheta$. Moreover, it holds that the mean of zeroth LLS is {\it always larger} than the mean level spacing $\Delta_N$; yet, it is the largest of all mean local spacings:
\begin{eqnarray}\label{FN-inequality}
    \!\!\!\! \langle s_0^{\rm loc}(\varphi;N)\rangle > {\rm max}_{1\le \ell \le N-1}\left\{
        \Delta_N, \langle s_\ell^{\rm loc}(\varphi;N)\rangle
    \right\}.
\end{eqnarray}
\end{subequations}

Owing to the BGS conjecture~\cite{BGS-1984}, the $N\rightarrow\infty$ descendants of Eqs.~(\ref{local-mean-N}) and (\ref{FN-inequality}) should apply to spectra of real `maximally chaotic' quantum systems. Defining the dimensionless mean of the $\ell$-th LLS
\begin{eqnarray} \label{loc-mean0infinity}
    \langle s_\ell^{\rm loc}\rangle =  \lim_{N\rightarrow \infty} \frac{1}{\Delta_N} \langle s_\ell^{\rm loc} (\varphi;N) \rangle,
\end{eqnarray}
where $\ell$ is kept fixed, and identifying the limit
\begin{eqnarray}\label{L-probs}
    E^{(\beta)} \left( \ell; \lambda \right) = \lim_{N\rightarrow\infty} E_N^{{\rm C}\beta{\rm E}}\left(\ell;\frac{2\pi\lambda}{N}\right)
\end{eqnarray}
with the probability that an interval of length $\lambda$ contains exactly $\ell$ points belonging to the unfolded spectrum of ${\rm C}\beta{\rm E}(\infty)$ [see discussion below Eq.~(\ref{CbetaE-m-JPDF})], we realize that the mean of $\ell$-th LLS on the unfolded energy scale is described by three distinguished, $\beta$-dependent sequences of {\it universal numbers}
\begin{subequations}\label{1st-infinite-N}
\begin{eqnarray} \label{local-mean-inf}
    \langle s_\ell^{\rm loc}\rangle = (1+\delta_{\ell,0}) \int_{0}^{\infty} d\lambda\, E^{(\beta)} \left( \ell; \lambda \right).
\end{eqnarray}
The theoretical values of $\langle s_\ell^{\rm loc}\rangle$ for $\beta=1, 2, 4$ are summarized in Table~\ref{local-means-table}. Mirroring the finite-$N$ conclusion [Eq.~(\ref{FN-inequality})], the mean of zeroth local spacing appears to be the largest local mean, yet it always exceeds unity:
\begin{eqnarray}\label{F-inf-inequality}
    \langle s_0^{\rm loc}\rangle > {\rm max}_{\ell \ge 1}\left\{
        1, \langle s_\ell^{\rm loc}\rangle
    \right\}.
\end{eqnarray}
\end{subequations}
{\it Equations~(\ref{local-mean-N}) and (\ref{local-mean-inf}), inequalities Eqs.~(\ref{FN-inequality}) and (\ref{F-inf-inequality}), and the three universal sequences
${\{ \langle s_\ell ^{\rm{loc}}\rangle\}}$ highlighted in Table~\ref{local-means-table} represent the first main result of this Letter.}

\begin{table}[t]
\footnotesize
		\begin{tabular}{ |l |l |c |c |c |c |c |}
			\hline
			\multicolumn{2}{ | c | }{ \footnotesize{Mean local spacings}} & $\langle s^\textrm{loc}_{0}\rangle$ & $\langle s^\textrm{loc}_{1} \rangle$ & $\langle s^\textrm{loc}_2 \rangle$ & $\langle s^\textrm{loc}_3 \rangle$ & $\langle s^\textrm{loc}_4 \rangle$ \\
			\hline\hline
			\multirow{2}{*}{\,$\beta=1$}& COE (theory) &  \bftab{1.28553}& \bftab{0.92267}& \bftab{0.97510}& \bftab{0.98856}& \bftab{0.99354}\\
			\cline{2-7}
			& Experiment & 1.28539& 0.92270& 0.97501& 0.98858& 0.99386\\
			\hline
			\multirow{2}{*}{\,$\beta=2$}& CUE (theory) & \bftab{1.17999}& \bftab{0.94449}& \bftab{0.98610}& \bftab{0.99404}& \bftab{0.99671}\\
			\cline{2-7}
			& Experiment & 1.17999& 0.94448& 0.98607& 0.99403& 0.99667\\
			\hline
			\multirow{2}{*}{\,$\beta=4$}& CSE (theory) & \bftab{1.10410}& \bftab{0.96536}& \bftab{0.99288}& \bftab{0.99702}& \bftab{0.99836}\\
			\cline{2-7}
			& Experiment & 1.10412& 0.96525& 0.99291& 0.99690& 0.99841\\
			\hline
			\hline
			\multirow{2}{*}{\,$\beta=0$\;}& Poisson (th) & \bftab{2\phantom{.00000}} & \bftab{1\phantom{.00000}} & \bftab{1\phantom{.00000}} & \bftab{1\phantom{.00000}} & \bftab{1\phantom{.00000}} \\
			\cline{2-7}
			& Experiment & 1.99994& 1.00019& 1.00006& 1.00020& 1.00000\\
			\hline
		\end{tabular}
		\caption{Comparison of theoretical and experimental values of mean LLS $\langle s^{\rm loc}_\ell\rangle$ in ${\rm C}\beta{\rm E}(N)$ and the Poisson ${(\beta=0)}$ spectra. Theoretical values for infinite-dimensional circular ensembles were extracted from Appendix~A of Ref.~\cite{B-2010}. Experimental values were produced from numerically generated ${\rm CUE}(N)$ ($2\times 10^8$ samples), ${\rm COE}(N)$, ${\rm CSE}(N)$ and Poisson ($2\times 10^7$ samples) ensembles with $N=1024$ and $\varphi=\pi$ ($\varphi=0$ for $\beta=0$).
		}
\label{local-means-table}
\end{table}

\textit{\textbf{Size-biased sampling and the inspection paradox}.}---Non-perturbative formulae for local means [Eqs.~(\ref{local-mean-N}) and (\ref{local-mean-inf})], proven in Appendix A, are explicit yet obscure: their appearances do not shed light onto the origin of nontrivial statistics of LLS. The
inequalities Eqs.~(\ref{FN-inequality}) and (\ref{F-inf-inequality}) are more helpful. They imply that a level spacing containing the `observation point' $\varphi$ (which is the zeroth LLS) is {\it stochastically larger} than the spacings between randomly chosen consecutive eigenlevels. This is the essence of the inspection paradox~\cite{MH-2020,F-1971,PKR-2022} in the probability theory. It occurs because the spacing {\it sampled locally} (around energy $\varphi$) is size-biased (the likelihood that $\varphi$ belongs to a chosen interval is proportional to its size) while the spacings between randomly chosen consecutive eigenlevels are not!

To make this claim quantitatively evident, let us consider a distribution $p_\ell(s;N)$ of the $\ell$-th LLS ${s_\ell^{\rm loc}(\varphi;N) = \theta_{n(\varphi)+\ell+1}-\theta_{n(\varphi)+\ell}}$ associated with a fixed reference point $\varphi$, see Fig.~\ref{Fig_Local_Spacings}. Since the fluctuational properties of $s_\ell^{\rm loc}(\varphi;N)=s_{n(\varphi)+\ell}$ in ${\rm C}\beta{\rm E}(N)$ spectra cannot depend on the position of a reference point, one may consider $\varphi$ to be random, chosen uniformly from the interval $[0,2\pi)$. It then follows that
\begin{eqnarray}\label{PL-der-01}
    p_\ell(s;N) = {\mathbb E}_{\bm{\theta}} {\mathbb E}_{{\rm \varphi}|{\bm \theta}} \big[
        \delta(s - s_\ell^{\rm loc}(\varphi;N))
    \big],
\end{eqnarray}
where ${\mathbb E}_{{\rm \varphi}|{\bm \theta}}$ denotes averaging with respect to a random choice of $\varphi$ for a given realization of the ${\rm C}\beta{\rm E}(N)$ spectrum. This inner mean in Eq.~(\ref{PL-der-01}) equals
\begin{eqnarray}\label{PL-der-02}
    {\mathbb E}_{{\rm \varphi}|{\bm \theta}} \big[ \delta(s -
        s_{n(\varphi)+\ell}
    \big]
     = \sum_{k=1}^{N} \delta(s - s_{k+\ell})\, P(\varphi \in s_k),
\end{eqnarray}
where $P(\varphi \in s_k)=s_k/2\pi$ is the probability that a randomly chosen reference point $\varphi$ falls into an arc connecting consecutive eigenlevels $\theta_k$ and $\theta_{k+1}$. {\it It is this probability that accounts for a length-dependent bias accompanying a local sampling of the spectrum}. Combining Eqs.~(\ref{PL-der-01}) and (\ref{PL-der-02}), we derive:
\begin{eqnarray}\label{PL-alt-02}
    p_\ell(s;N) = \frac{1}{\Delta_N}\, {\mathbb E}_{\bm{\theta}}\, {\mathbb E}_{X}
    \left[
        s_X\, \delta(s - s_{X+\ell})
    \right].
\end{eqnarray}
Here $\Delta_N=2\pi/N$ is the mean spacing between consecutive eigenlevels and $X$ is a uniformly distributed random variable taking the values $\{1, 2, \dots, N\}$. The bias becomes even more transparent in the distribution of zeroth LLS~\cite{Rem-L,PRK-TA}. Indeed, setting $\ell=0$ in Eq.~(\ref{PL-alt-02}) and consulting Eq.~(\ref{PN-consecutive}), we observe a suggestive factorization of $p_0(s;N)$ into a product of the traditional level spacing distribution $p_N^{{\rm C}\beta{\rm E}}(s)$ and the weight $s/\Delta_N$ of a length-dependent bias (see Supplemental Material~\cite{TRK-2023-sm} for further details).

Equation~(\ref{PL-alt-02}) supplies the mean of the $\ell$-th LLS:
\begin{eqnarray} \label{PL-alt-01}
   \langle s_\ell^{\rm loc}(0;N)\rangle = \frac{1}{\Delta_N}\, {\mathbb E}_{\bm{\theta}}\, {\mathbb E}_{X}
    \left[
        s_X\, s_{X+\ell}
    \right].
\end{eqnarray}
This representation has several important consequences. (i) First, Eq.~(\ref{PL-alt-01}) makes it manifestly evident that means of local spacings are intrinsically biased through an extra factor $s_X/\Delta_N$; it is precisely this biasing that distorts the statistics in counter-intuitive ways. (ii) Second, Eq.~(\ref{PL-alt-01}) implies the inequalities Eqs.~(\ref{FN-inequality}) and (\ref{F-inf-inequality}) which underlined our earlier discussion of the inspection paradox in the random-matrix-theory setting; their proofs are given in the Supplemental Material~\cite{TRK-2023-sm}. (iii) Third, Eq.~(\ref{PL-alt-01}) indicates that a measurement of the average of $\ell$-th local spacing provides a {\it direct access} to the auto-covariance~\cite{M-2004,RTK-2023} of level spacings located $\ell$ eigenlevels apart:
\begin{eqnarray}\label{cov-matrix}
    {\rm cov}_{{\bm \theta}, X} [s_X, s_{X+\ell}] = \Delta_N^2 \left(\frac{ \langle s_\ell^{\rm loc}(0;N)\rangle}{\Delta_N}-1\right).
\end{eqnarray}
(iv) Finally, we notice that both Eq.~(\ref{PL-alt-01}) and its obvious $N\rightarrow\infty$ counterpart, must be equivalent to Eqs.~(\ref{local-mean-N}) and (\ref{local-mean-inf}), respectively.

The same mechanism of a size-biased sampling is at work in the Poisson spectra ${(\beta=0)}$. In this case Eq.~(\ref{local-mean-inf}) stays valid provided~\cite{F-1971}
$ E^{(0)} \left( \ell; \lambda \right) = \lambda^\ell e^{-\lambda}/\ell!$. Equation (\ref{local-mean-inf}) immediately supplies yet another counter-intuitive result: $\langle s_0^{\rm loc}\rangle =2$ while $\langle s_\ell^{\rm loc}\rangle =1$ for all ${\ell\ge 1}$. This corresponds to a famous example of the inspection paradox in the Poisson point process: the average waiting time for a bus by a person which arrives at a bus station at some random uniformly distributed time is twice as large as a na\"ive expectation~\cite{IP-Remark} given by half of the average time between consecutive buses~\cite{MH-2020,F-1971}.

In the context of quantum chaology, in view of the Berry-Tabor conjecture~\cite{BT-1977}, this implies that the universal sequence $\{2, 1, 1, \dots \}$ for LLS means $\{ \langle s_0^{\rm loc}\rangle, \langle s_1^{\rm loc}\rangle , \langle s_2^{\rm loc}\rangle , \dots\}$ should be observable in the unfolded spectra of generic quantum systems with integrable classical dynamics.

\textit{\textbf{Theory vs numerical experiments: unfolded and raw spectra.}}---Let us confront the universal predictions for LLS means $\langle s_\ell^{\rm loc}\rangle$ with the results of numerical experiments performed for a variety of random matrix models and systems belonging to the Wigner-Dyson and Poisson universality classes. Two different numerical protocols (see Protocols 1 and 2.1/2.2 detailed in Appendix B) should be employed for statistical analysis of local level spacings in random and deterministic systems.

\textit{\textbf{Test~I:~Circular ensembles vs Poisson sequences.}}---Table~\ref{local-means-table} summarizes results of numerical experiments performed, within Protocol 1, for random spectra of both the Dyson triad ${\rm C}\beta{\rm E}(N)$ and the Poisson spectral sequences. In all cases, universal theoretical values of mean local spacings agree well with the numerics: they have been checked to lie inside 99\% confidence intervals around numerically evaluated local means.

\begin{table}[b]
\footnotesize
\begin{tabular}{| l| l |l |l |l |l |l |}
\hline
      \multicolumn{2}{ | c | }{\footnotesize{Mean local spacings}}
& $\; \,\langle s^\textrm{loc}_0\rangle$  & $\;\,\langle s^\textrm{loc}_1 \rangle$ &
              $\;\, \langle s^\textrm{loc}_2 \rangle$ & $\;\, \langle s^\textrm{loc}_3 \rangle$ & $\;\, \langle s^\textrm{loc}_4 \rangle$ \\
\hline \hline
$\beta=2$& Riemann zeros & 1.17846& 0.94363& 0.98568& 0.99414& 0.99651 \\
\hline \hline
$\beta=0$& Rect. billiard & 1.99812 & 1.00172 & 0.99986 & 1.00004 & 0.99960 \\
            \hline
\end{tabular}
\caption{
Experimental values of mean LLS $\langle s^{\rm loc}_\ell\rangle$ for unfolded (i) zeros of the Riemann zeta function and (ii) eigenlevels of a quantum rectangular billiard with irrational squared aspect ratios. For theoretical values, see Table~\ref{local-means-table}.}\label{table-Riemann-billiard}
\end{table}

{\it Since the four universal theoretical sequences, presented in Table~\ref{local-means-table}, are unique and clearly distinguish between the Wigner-Dyson ($\beta=1, 2, 4$) and Poisson ($\beta=0$) universality classes, the LLS means can be employed to uncover underlying classical dynamics of quantum systems.}

\begin{table}[t]
\footnotesize
		\begin{tabular}{ |l |l |c |c |c |}
			\hline
			\multicolumn{2}{ | c | }{ Ratios of local means} & $\langle s^\textrm{loc}_1 \rangle / \langle s^\textrm{loc}_0 \rangle $ & $\langle s^\textrm{loc}_2 \rangle / \langle s^\textrm{loc}_0 \rangle $ & $\langle s^\textrm{loc}_3 \rangle / \langle s^\textrm{loc}_0 \rangle $ \\
			\hline\hline
			\multirow{2}{*}{$\;\beta=1\;$}& ${\rm COE}$ (theory) & \bftab{0.71773}& \bftab{0.75852}& \bftab{0.76899}\\
			\cline{2-5}
			& ${\rm SYK}_4$ $({\mathcal N}=24)$& 0.71794& 0.75856& 0.76887\\
			\hline
			\hline
			\multirow{2}{*}{$\;\beta=2\;$}& ${\rm CUE}$ (theory) & \bftab{0.80042}& \bftab{0.83569}& \bftab{0.84241}\\
			\cline{2-5}
			& ${\rm SYK}_4$ $({\mathcal N}=26)$& 0.80047& 0.83613& 0.84277\\
			\hline
			\hline
			\multirow{2}{*}{$\;\beta=4\;$}& ${\rm CSE}$ (theory) & \bftab{0.87434}& \bftab{0.89927}& \bftab{0.90301}\\
			\cline{2-5}
			& ${\rm SYK}_4$ $({\mathcal N}=28)$ & 0.87431& 0.89947& 0.90291\\
			\hline\hline
            \multirow{1}{*}{$\;\beta=0\;$}& Poisson (th)${}_{\phantom{0_0}}$ & \bftab{1/2} & \bftab{1/2} & \bftab{1/2}\\
            \hline
            		\end{tabular}
		\caption{Experimental values for ratios of LLS means in the raw spectra of ${\rm SYK}_4$ models with $J=4$. They were produced by numerical diagonalization of ${\rm SYK}_4$ Hamiltonians [Eq.~(\ref{SYK-Ham})]; $10^6$ samples with the same reference point $\varphi=0$ were used. For comparison, we specified the universal theoretical values of ratios computed with the help of Table~\ref{local-means-table}.
}
        \label{table-SYK-ratios}
\end{table}

\textit{\textbf{Test~II:~Riemann zeta zeros vs integrable billiards.}}---This observation is further confirmed for two paradigmatic deterministic systems of quantum chaology -- the Riemann zeta function and an irrational rectangular billiard. Table~\ref{table-Riemann-billiard} presents the values of mean LLS obtained by {\it statistical} analysis of these two systems. For the Riemann zeta function, Protocol~2.1 with $Q=10^6$ was applied to Odlyzko's data set of $10$ billion zeros located around $10^{23}$-rd zero~\cite{O-set,HO-2011,O-2001}. For rectangular billiards, we used Protocol~2.2 with $\varphi=10^{12}$ and parameter $h$ being the billiard aspect ratio whose variation does not affect the mean level density at high energies~\cite{RK-2021} provided the billiard area is kept fixed. As was expected, the LLS means, computed on the basis of experimental data~\cite{Rem-Riemann}, unequivocally identify a spectral universality class each of the two systems belongs to.

\textit{\textbf{Test~III:~Quantum~many-body~systems -- proof of concept.}}---Statistical tests performed so far referred to spectral data obtained from {\it unfolded} spectra. Remarkably, the effect of locality in level spacing fluctuations is robust enough to be clearly observed in the raw spectra by studying the {\it ratios of mean local spacings} (which is different from the Oganesyan-Huse-Bogomolny $r$-statistics~\cite{OH-2007,ABGR-2013} dealing with the {\it mean of ratios} between {\it consecutive spacings}).

To be specific, we focus on the Sachdev-Ye-Kitaev (${\rm SYK}_q$) model which has become a paradigm of quantum many-body physics~\cite{SY-1993,K-2015,Rem-SYK}. For $q=4$, it describes ${\mathcal N}$ Majorana fermions subject to a random, infinite-range, four-body interaction
\begin{eqnarray}\label{SYK-Ham}
    \!\! H_{{\rm SYK}_4} = \sum_{1\le i_1 < i_2 < i_3 < i_4 \le {\mathcal N}} J_{i_1 i_2 i_3 i_4} \chi_{i_1} \chi_{i_2}\chi_{i_3}\chi_{i_4},
\end{eqnarray}
where $\chi_j$ are Majorana fermions satisfying the Clifford algebra $\{\chi_j,\chi_k\}=\delta_{jk}$, and $J_{i_1 i_2 i_3 i_4}$ are independent real Gaussian variables with zero mean and variance $6J^2/{\mathcal N}^3$.

As the mean level density in this model depends exponentially on the energy~\cite{GGV-2017}, it serves as a showcase for testing effects of locality in the raw spectra. In
Table~\ref{table-SYK-ratios} we have summarized the results of numerical simulations for the {\it ratios} of {\it means of local spacings}
\begin{eqnarray}
    \varrho_\ell^{\rm loc}=\frac{\langle s_\ell^{\rm loc}\rangle}{\langle s_0^{\rm loc}\rangle},
\end{eqnarray}
calculated for various $\ell\ge 1$. (The LLS means were computed by applying Protocol 1 to the {\it raw} ${\rm SYK}_4$ spectra). This measure is a natural choice since a ratio of local means is barely affected by a system dependent mean level density~\cite{T-III-conjecture,AMR-2022}. Being well-defined for both Wigner-Dyson and Poissonian spectra, the ratio $\varrho_\ell^{\rm loc}$ satisfies the inequality ${0 \le \varrho_\ell^{\rm loc} \le 1}$, see Eq.~(\ref{F-inf-inequality}).

Comparison of theoretical and experimental values of $\varrho_\ell^{\rm loc}$ clearly indicates that this local statistics, applied to the raw spectra, does uncover the universal aspects of spectral fluctuations, placing random ${\rm SYK}_4$ Hamiltonians into the right (Wigner-Dyson) universality class with the symmetry index suggested by the Bott periodicity in number of Majorana fermions~\cite{YLX-2017}. {\it This proof of concept is the second main result of the Letter.}

\textit{\textbf{Acknowledgements.}}---The authors thank A.~M. Garc\'ia-Garc\'ia for the correspondence on recursion relations for representation matrices of Majorana fermions in
the ${\rm SYK}_4$ model. F.~Bornemann is thanked for providing us with the MATLAB package for numerical evaluation of Fredholm determinants. Last but not least, we are grateful to A.~M.~Odlyzko for sharing with us the Riemann zeros data set used in this research. This work was supported by the Israel Science Foundation through the Grant No. 428/18. Some of the computations presented in this work were performed on the Hive computer cluster at the University of Haifa, which is partially funded through the Israel Science Foundation Grant No. 2155/15.

\textit{\textbf{Appendix A: Proof of Eqs.~(\ref{local-mean-N}) and (\ref{local-mean-inf}).}}---A formal proof of Eq.~(\ref{local-mean-N}) is based on the integral identity ($z \in {\mathbb C}$)
\begin{eqnarray}\label{proof-1}
    \!\!\!\!\!\!\int_{0}^{2\pi} d\vartheta\, z^{n(\vartheta)} = z^N s_0^{\rm loc}(0;N)  + \sum_{\ell=1}^{N-1} z^\ell s_\ell^{\rm loc}(0;N)
\end{eqnarray}
that expresses the generating function of local spacings $\{s_\ell^{\rm loc}(0;N)\}$ in its r.h.s. in terms of a random, integer-valued function $n(\vartheta)$ returning the index of the {\it left}-nearest-to-$\vartheta$ eigen-angle for all $\vartheta \in [0,2\pi)$, see Fig.~\ref{Fig_Local_Spacings} and related discussion. To assess {\it mean local spacings}, we average Eq.~(\ref{proof-1}) with respect to the random ${\rm C}\beta{\rm E}(N)$ spectrum. Introducing the eigen-angle counting function~\cite{M-2004} ${\mathcal N}(\vartheta)$ which equals the number of eigen-angles in the spectral interval $(0,\vartheta)$, and spotting that $n(\vartheta)={\mathcal N}(\vartheta)$ if ${\mathcal N}(\vartheta)>0$ and $n(\vartheta)=N$ if ${\mathcal N}(\vartheta)=0$, we derive:
\begin{eqnarray}\label{proof-2}
    \!\!\!{\mathbb E}_{\bm{\theta}} \big[ z^{n(\vartheta)}\big] &=& z^N \big[
        E_N^{{\rm C}\beta{\rm E}}(0;\vartheta) + E_N^{{\rm C}\beta{\rm E}}(0;2\pi-\vartheta)
    \big] \nonumber\\
     &+& \sum_{\ell=1}^{N-1} z^\ell E_N^{{\rm C}\beta{\rm E}}(\ell;\vartheta).
\end{eqnarray}
Here, $E_N^{{\rm C}\beta{\rm E}}(\ell;\vartheta)={\rm Prob}({\mathcal N}(\vartheta)=\ell)$ is the probability to observe exactly $\ell$ eigen-angles in an arc of length $\vartheta$. Substituting Eq.~(\ref{proof-2}) back to the averaged Eq.~(\ref{proof-1}), we reproduce the sought Eq.~(\ref{local-mean-N}). The infinite-dimensional version [Eq.~(\ref{local-mean-inf})] of this result follows upon implementing the $N\rightarrow\infty$ limit defined by Eqs.~(\ref{loc-mean0infinity}) and (\ref{L-probs}).

\textit{\textbf{Appendix B: Numerical protocols for LLS in random and deterministic systems.}}---Throughout the Letter, statistical analysis of LLS in random and deterministic systems is performed within two (properly adjusted) general protocols.

\textit{\textbf{Protocol 1}}. For systems with intrinsic randomness, we choose a pre-defined reference point $\varphi$ in the unfolded spectrum, record $M$ realizations of local spacings $\{ \{ s_\ell^{{\rm loc}\,(1)}(\varphi)\},\dots, \{ s_\ell^{{\rm loc}\,(M)}(\varphi)\}\}$, and further perform sample averaging. This protocol was used in Test~I.

\textit{\textbf{Protocol 2}}. For deterministic systems, an artificial randomization should be introduced first. To this end, one may either choose a set of $Q$
random reference points $\{ \varphi_1,\dots, \varphi_Q \}$ in the unfolded spectrum or randomize a suitable intrinsic system parameter, denoted $h$, such that its variation does not affect the mean spectral density. Having recorded, out of the unfolded spectrum, $Q$ sets of local spacings $\{ \{s_\ell^{\rm loc}(\varphi_1)\},\dots, \{s_\ell^{\rm loc}(\varphi_Q)\}\}$ (Protocol~2.1) or $\{ \{ s_\ell^{\rm loc}(\varphi; h_1)\},\dots, \{ s_\ell^{\rm loc}(\varphi; h_Q)\}$, where $\varphi$ is fixed while $\{ h_1,\dots, h_Q\}$ are chosen at random (Protocol~2.2), we perform sample averaging for each $\ell$ of interest. Both Protocols were used in Test~II.

\newpage
\onecolumngrid
\section{Supplemental Material}
\beginsupplement
\renewcommand{\theequation}{S.\arabic{equation}}
\setcounter{equation}{0}
\onecolumngrid
\supplsection{Inequalities for the means of local spacings}
\noindent
Below we shall prove two inequalities combined into the single statement Eq.~(\ref{FN-inequality}) of the main text.
\noindent\newline\newline
{\bf Proof of the first inequality}.---To prove the first inequality
\begin{eqnarray} \label{ineq-01}
    \langle s_0^{\rm loc}(\varphi;N) \rangle > \Delta_N,
\end{eqnarray}
we make use of Eq.~(\ref{PL-alt-01}) taken at $\ell=0$ to write down
\begin{equation}\label{ineq-01-aux}
    \langle s_0^{\rm loc}(\varphi;N) \rangle = \frac{1}{\Delta_N} {\mathbb E} \left[s_X^2\right],
\end{equation}
where ${\mathbb E}[\dots] = {\mathbb E}_{{\bm\theta}}{\mathbb E}_X[\dots]$. Since the variance of the random variable $s_X$ is always positive, the inequality ${{\mathbb E} \left[s_X^2\right] > \left({\mathbb E} \left[s_X\right]\right)^2}$ holds, so that
\begin{eqnarray}
    \langle s_0^{\rm loc}(\varphi;N) \rangle > \frac{1}{\Delta_N} \left({\mathbb E} \left[s_X\right]\right)^2 = \Delta_N.
\end{eqnarray}
This ends the proof of Eq.~(\ref{ineq-01}).
\noindent\newline\newline
{\bf Proof of the second inequality}.---To prove the second inequality
\begin{eqnarray}\label{ineq-02}
    \langle s_0^{\rm loc}(\varphi;N) \rangle > \langle s_\ell^{\rm loc}(\varphi;N) \rangle
\end{eqnarray}
which holds for all $\ell=1, 2, \dots, N-1$, we start with the representation
\begin{eqnarray}\label{ineq-02-aux}
    \langle s_\ell^{\rm loc}(\varphi;N) \rangle = \frac{1}{\Delta_N} {\mathbb E} \left[s_X s_{X+\ell}\right],
\end{eqnarray}
see Eq.~(\ref{PL-alt-01}), and notice that ${\mathbb E} \left[(s_X - s_{X+\ell})^2\right]>0$ for all $\ell=1,\dots,N-1$. This inequality, combined with Eq.~(\ref{ineq-02-aux}), yields
\begin{eqnarray}
\!\!\!\!\!\!\!
    \langle s_\ell^{\rm loc}(\varphi;N) \rangle < \frac{1}{2} \left( \frac{1}{\Delta_N} {\mathbb E} \left[s_X^2\right] +   \frac{1}{\Delta_N} {\mathbb E} \left[s_{X+\ell}^2\right]\right).
\end{eqnarray}
Realizing that both terms in the brackets are equal to each other and invoking Eq.~(\ref{ineq-01-aux}), we reproduce the sought inequality Eq.~(\ref{ineq-02}).

\begin{figure}[t]
\includegraphics[width=0.75\columnwidth]{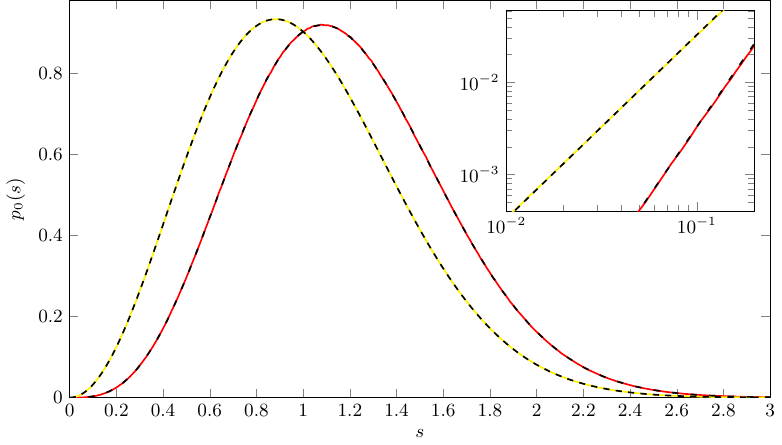}
\caption{Comparison of theoretical and experimental distributions of the zeroth local level spacings for the {\rm CUE} spectra. Yellow and red curve display experimental curves $p(s)$ -- the distribution of spacings between consecutive eigenlevels and $p_0(s)$ -- the distribution of zeroth local spacings, respectively. They were obtained from $2\times10^8$ samples with $N=1024$ and $\varphi=\pi$. Dashed curves appearing on top of experimental curves show corresponding theoretical $N\rightarrow\infty$ laws given by Eqs.~(\ref{p-beta}), (\ref{E2}) and (\ref{painleve-5-all}) for $p(s)$, Eq.~(\ref{PL0-infinity-sm}) for $p_0(s)$. The inset shows magnified experimental distributions at the origin plotted on the log-log scale. The dashed lines indicate power law dependencies with the slope $2$ and $3$.} \label{Fig-CUE-local}
\end{figure}

\supplsection{Distribution of the zeroth local spacing}
\noindent

In the main text, below Eq.~(\ref{PL-alt-02}), we have derived a simple and suggestive formula for the distribution of zeroth local spacing. In the ${\rm C}\beta{\rm E}(N)$ setting, one has
\begin{eqnarray}\label{PL0-N-sm}
p_0(s;N) = \frac{s}{\Delta_N} \, p_N^{{\rm C}\beta{\rm E}}(s),
\end{eqnarray}
where $\Delta_N=2\pi/N$, and $p_N^{{\rm C}\beta{\rm E}}(s)$ is the traditional level spacing distribution, see Eq.~(\ref{PN-consecutive}) of the main text for the definition. The physically motivated $N\rightarrow\infty$ scaling limit
\begin{eqnarray} \label{N-infinity-lim-sm}
    p_0^{(\beta)}(s) = \lim_{N\rightarrow\infty} \frac{2\pi}{N} p_0\left(\frac{2\pi}{N}s;N\right)
\end{eqnarray}
produces three universal distributions
\begin{eqnarray}\label{PL0-infinity-sm}
p_0^{(\beta)}(s) = s p^{(\beta)}(s).
\end{eqnarray}
Marked by the Dyson index $\beta$, they are directly related to the Wigner-Dyson level spacing distribution $p^{(\beta)}(s)$; their explicit forms required for numerical tests are given by Eqs.~(\ref{p-beta})--(\ref{PV-bc}) below.

Notice that the multiplicative factor $s$ appearing in Eqs.~(\ref{PL0-N-sm}) and (\ref{PL0-infinity-sm}) modifies an `effective repulsion' between the eigenlevels which happened to crown the `inspected' spacing, making this local repulsion stronger: ${p_0^{(\beta)}(s) = c_\beta s^{\beta+1}+{\mathcal O}(s^{\beta+2})}$ instead of the usual Wigner-Dyson repulsion ${p^{(\beta)}(s) = c_\beta s^\beta +{\mathcal O}(s^{\beta+1})}$ as $s\rightarrow 0$, where $c_\beta$ is a known constant. Our numerical tests unequivocally support this conclusion.

\begin{figure}[t]
\includegraphics[width=0.75\columnwidth]{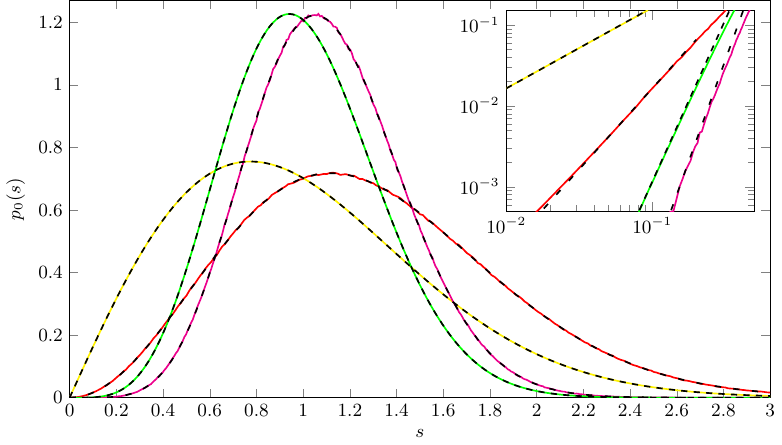}
\caption{
Distribution of zeroth local level spacing for the ${\rm COE}$ and ${\rm CSE}$ spectra. For the ${\rm COE}$ spectra: red and yellow curves show experimental distribution $p_0(s)$ and $p(s)$, respectively. For the ${\rm CSE}$ spectra: magenta and green curves display experimental distributions $p_0(s)$ and $p(s)$, respectively. The curves were produced by statistical analysis of $2\times10^7$ samples with $N=1024$ and $\varphi=\pi$. The dashed lines on top of experimental curves show theoretical predictions for $p_0(s)$ and $p(s)$ determined by Eqs.~(\ref{PL0-infinity-sm}) and (\ref{p-beta})--(\ref{painleve-5-all}). The inset shows magnified experimental distributions at the origin plotted on the log-log scale. The dashed lines there indicate the predicted power law dependencies with the slopes $1$, $2$, $4$ and $5$ (from left to right).
}\label{Fig-COE-CSE-local}
\end{figure}
\noindent\newline
{\bf Discussion of numerical tests}.---In Fig.~\ref{Fig-CUE-local} we compare the distribution of zeroth local spacing for numerically simulated large-dimensional ${\rm CUE}(N)$ spectra with the theoretical prediction for $p_0(s)$. The agreement is perfect. For $s\ll 1$, the distribution $p_0(s)$ shows a cubic slope, in concert with the remark below Eq.~(\ref{PL0-infinity-sm}).

\begin{figure}[t]
\includegraphics[width=0.75\columnwidth]{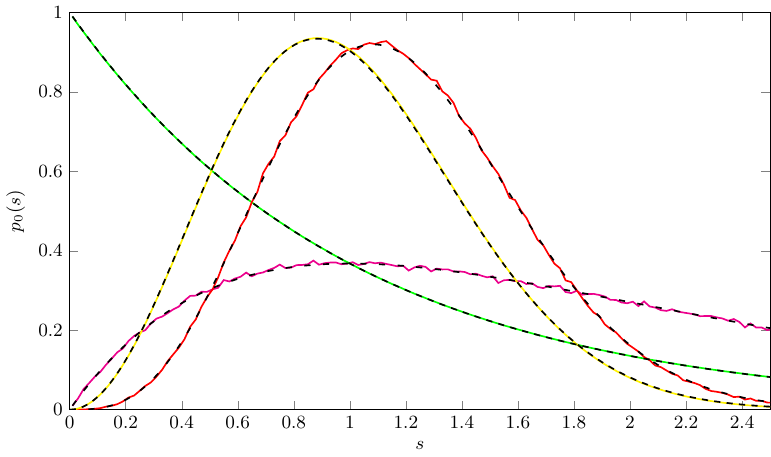}
\caption{Comparison of theoretical and experimental distributions of zeroth local level spacings for zeros of the Riemann zeta function and spectra of irrational rectangular billiards. For the Riemann zeta function: red and yellow curves represent the experimentally calculated local spacing distribution $p_0(s)$ and the distribution $p(s)$ between consecutive zeros, respectively. For rectangular billiards: magenta and green curves display experimentally determined local spacing distribution $p_0(s)$ and the traditional spacing distribution $p(s)$, respectively. Statistical analysis of both systems involved $10^6$ samples (see description of Test II in the main text for more details). The dashed lines show theoretical predictions for $p(s)$ and $p_0(s)$. Theoretical curves for the Riemann zeta function coincide with those displayed in Fig.~\ref{Fig-CUE-local}. Theoretical curves for the billiard spectra are given $p(s)= e^{-s}$ and $p_0(s) = s e^{-s}$ corresponding to the Poisson spectra.}\label{Fig-Riemann-billiard}
\end{figure}

In Fig.~\ref{Fig-COE-CSE-local}, we display the experimentally computed distribution of zeroth local spacing for large-dimensional ${\rm COE}(N)$ and ${\rm CSE}(N)$ spectra. For comparison with the theory and details of statistical analysis, the reader is referred to the caption.

In Fig.~\ref{Fig-Riemann-billiard}, our theoretical prediction is confronted with experimental distribution of zeroth local spacing determined for two real systems: the Riemann zeta function and irrational rectangular billiards.

The experimental curves obtained for nontrivial zeros of the Riemann zeta function follow closely the prediction derived for the ${\rm CUE}$ spectra. In distinction to the experimental ${\rm CUE}$ curve displayed in Fig.~\ref{Fig-CUE-local}, the fluctuations in Fig.~\ref{Fig-Riemann-billiard} are more pronounced. This is hardly surprising since $200$ times less samples were produced out of available Odlyzko's sets for nontrivial Riemann zeros.

The experimental curves for rectangular billiards are fundamentally different; they nicely follow the theoretical predictions for the Poisson spectra specified in the figure caption. Even though there is no level repulsion in the Poisson spectra, the distribution of zeroth local spacing exhibits an {\it effective} linear repulsion between the eigenlevels located at the endpoints of the `inspected' (zeroth) spacing.
\noindent\newline\newline
{\bf Distribution of spacings between consecutive eigenlevels and the fifth Painlev\'e transcendent.}---For comparison of experimental and theoretical results for the distribution $p_0^{(\beta)}(s)$ of zeroth local spacing [Eq.~(\ref{PL0-infinity-sm})], we have used the following nonperturbative formulae for the distributions $p^{(\beta)}(s)$ of consecutive spacings ($\beta=1,2,4$):
\begin{eqnarray}\label{p-beta}
    p^{(\beta)}(s) = \frac{d^2}{ds^2} E^{(\beta)}(0;s),
\end{eqnarray}
where~\cite{FW-2000-sm,F-2005-sm}
\begin{subequations}
\begin{eqnarray} \label{E2}
    E^{(2)}(0;s) =\exp\left(
    \int_{0}^{2\pi s} \frac{\sigma_0(t)}{t} dt
    \right),
\end{eqnarray}
\begin{eqnarray}\label{E-1}
    E^{(1)}(0;s) =\exp\left(
    -\frac{1}{2}\int_{0}^{2\pi s} \sqrt{-\frac{d}{dt}\frac{\sigma_0(t)}{t}} dt
    \right)\, \sqrt{E^{(2)}(0;s)}
\end{eqnarray}
and
\begin{eqnarray}\label{E4}
    E^{(4)}(0;s/2) =\cosh\left(
    \frac{1}{2}\int_{0}^{2\pi s} \sqrt{-\frac{d}{dt}\frac{\sigma_0(t)}{t}} dt
    \right)\, \sqrt{E^{(2)}(0;s)}
\end{eqnarray}
\end{subequations}
are the gap formation probabilities $E^{(\beta)}(0;s)$ for the ${\rm Sine}_\beta$ point process. Here, $\sigma_0(t)$ is a fifth Painl\'eve transcendent satisfying the nonlinear differential equation
\begin{subequations}\label{painleve-5-all}
\begin{eqnarray}\label{PV}
    (t \sigma_0^{\prime\prime})^2 +  (t\sigma_0^\prime-\sigma_0) \left(t\sigma_0^\prime -\sigma_0 + 4 (\sigma_0^\prime)^2\right) = 0
\end{eqnarray}
subject to the boundary condition
\begin{eqnarray}\label{PV-bc}
    \sigma_0(t) = -\frac{t}{2\pi} - \left(\frac{t}{2\pi}\right)^2 + {\mathcal O}(t^3)
\end{eqnarray}
\end{subequations}
as $t\rightarrow 0$. Notice that, contrary to the phenomenological Wigner-surmise formulae, the above representations are {\it exact}.

A fifth Painlev\'e transcendent $\sigma_0(t)$ [Eq.~(\ref{painleve-5-all})] as well as the integrals appearing in Eqs.~(\ref{E2})--(\ref{E4}) were determined numerically by employing a standard MATLAB ordinary differential equations (ODE) solver applied to the Chazy form of Eq.~(\ref{PV}), see Eq.~(B.7) of Ref.~\cite{RK-2023-sm}. To produce initial conditions for the ODE solver away from the singularity at $t=0$, we used a Taylor polynomial of a high degree generated analytically through the recurrence relation Eq.~(B.8) of Ref.~\cite{RK-2023-sm} after setting $\zeta=1$ therein.
\noindent\newline\newline
\supplsection{On improving statistics for ratios $\varrho_\ell^{\rm loc}$ of means of local spacings}

The statistical analysis of the ratios $\varrho_\ell^{\rm loc}$ in Test III was based on $10^6$ samples of {\it raw} spectra with a single reference point $\varphi=0$. Let us stress that statistics of comparable quality for $\varrho_\ell^{\rm loc}$ could be obtained with a significantly smaller number ($M$) of samples. Indeed, choosing either deterministically or at random a sufficiently large set of pre-defined local reference points $\{\varphi_1, \varphi_2,\dots,\varphi_Q\}$, one could first evaluate sample averages $\{ \langle s_\ell^{\rm loc}(\varphi_\alpha)\rangle_M \}|_{\alpha=1}^Q$ of local spacings separately for each $\varphi_\alpha$ and then use them to calculate a set of local ratios
\begin{equation}
\rho^\textrm{loc}_\ell(\varphi_\alpha)=\frac{\langle s_\ell^\textrm{loc}(\varphi_\alpha) \rangle_M}{\langle s_0^\textrm{loc}(\varphi_\alpha) \rangle_M}
\end{equation}
for each $\ell$ and $\alpha=1,2,\dots, Q$. Since the theoretical expectation values of these ratios should not depend on a particular value of $\varphi_\alpha$, one may further average them over $\alpha$
\begin{eqnarray}
     \varrho^{\rm loc}_\ell=\frac{1}{Q}\sum_{\alpha=1}^Q \rho^{\rm loc}_\ell(\varphi_\alpha)
\end{eqnarray}
to improve the statistics. The same strategy can be used to improve statistics of local spacings in the {\it unfolded} spectra, see Test I.
\begin{figure}[t]
\includegraphics[width=0.4\columnwidth]{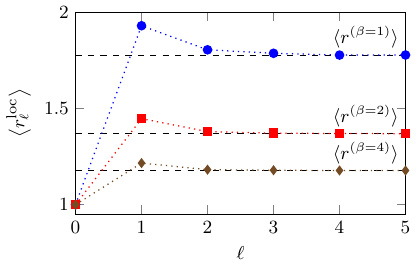}
\caption{The mean values $\langle r_\ell^{\rm loc}\rangle$ of the local $r$-ratios calculated for large-dimensional ${\rm COE}(N)$ (blue dots), ${\rm CUE}(N)$ (red squares) and ${\rm CSE}(N)$ (brown diamonds) matrix models. For simulation parameters, see caption to Table~\ref{local-means-table}. The values $\langle r^{(\beta)}\rangle$, represented by dashed lines, are approximately equal~\cite{ABGR-2013-sm} to $1.7781$ for $\beta=1$, $1.3684$ for $\beta=2$, and $1.1769$ for $\beta=4$.}\label{local-ratios-vs-bog}
\end{figure}

\supplsection{On a local version of Oganesyan-Huse-Bogomolny $r$-ratio}
A notion of local level spacings introduced in the Letter raises a natural question about a possible relation between the Oganesyan-Huse-Bogomolny $r$-ratio~\cite{OH-2007-sm,ABGR-2013-sm} and its local version
\begin{eqnarray}
    r_\ell^{{\rm loc}} = \frac{s_{\ell+1}^{{\rm loc}}}{s_\ell^{{\rm loc}}},
\end{eqnarray}
where $s_\ell^{{\rm loc}}$ is the (fluctuating) $\ell$-th local spacing with respect to the reference point $\varphi$. Similarly to the $r$-statistics, this ratio is also barely affected by a system dependent mean level density. Yet, contrary to the $r$-ratio, the average $\langle r_\ell^{\rm loc} \rangle$ is not a constant anymore, being $\ell$-dependent. More precisely, it is described by universal sequences $\{\langle r_\ell^{\rm loc} \rangle\}$ which depend on the spectral universality class and system symmetry; the first member of these sequences is always unity~\cite{Rem-unity}
\begin{eqnarray}\label{r0unity}
\langle r_0^{\rm loc}\rangle =1.
\end{eqnarray}
A truly nonperturbative calculation of $\langle r_\ell^{\rm loc}\rangle$ for $\ell \neq 0$ is a nontrivial problem.

In Fig.~\ref{local-ratios-vs-bog} we show the results of numerical simulations of local $r$-ratios, performed for large-dimensional ${\rm C}\beta{\rm E}(N)$ matrix models. The graphs suggest that, as $\ell$ grows, the universal sequences $\langle r_\ell^{\rm loc}\rangle$ start to approach the universal values $\langle r^{(\beta)} \rangle$ due to Bogomolny and co-authors~\cite{ABGR-2013-sm}. This is unsurprising since the effects of locality fade away as $\ell$ increases.

In the Poisson spectra, characterized by completely uncorrelated consecutive spacings, the memory of locality is lost immediately. Indeed, explicit calculation of the distribution functions $ p_\ell(r)$ of local $r$-ratios yields
\begin{eqnarray}
    p_\ell(r) = {\mathbb E}_{{\bm \theta}} \left[\delta\left(r - r_\ell^{\rm loc}\right)\right] = \left\{\begin{array}{cc}
                  \displaystyle \frac{2}{(1+r)^3}, & \ell=0; \\
                  \displaystyle \frac{1}{(1+r)^2}, & \ell \ge 1.
                \end{array}\right.
\end{eqnarray}
Hence, starting with $\ell=1$, the fluctuations of local ratios become indistinguishable from the Oganesyan-Huse-Bogomolny $r$-ratio whose distribution equals~\cite{ABGR-2013-sm}
\begin{eqnarray}
    p(r) = \frac{1}{(1+r)^2}.
\end{eqnarray}


\begin{references}
\bibitem{Rem-ratio} Since the mean of such a ratio diverges for the Poisson spectra, one actually deals with a minimum between a ratio of spacings and its inverse.
\bibitem{Rem-moments} The local spacings $s_\ell^{\rm loc}(\varphi;N)$ and $s_{N\pm \ell}^{\rm loc}(\varphi;N)$ have the same distributions due to the $N$-periodicity and the mirror symmetry.
\bibitem{Rem-phi-independent} Notice that the answer does not depend on $\varphi$.
\bibitem{Rem-L} The entire set $\{ p_\ell(s;N)\}$ of local spacing distributions for $\ell>0$ can be determined non-perturbatively as well~\cite{PRK-TA}.
\bibitem{IP-Remark} In particular, it completely misses the observation that a passenger is more likely to arrive between two busses that are far apart.
\bibitem{Rem-Riemann} We remark that experimental values of $\langle s_0^{\rm loc}\rangle$ computed for the Riemann zeros lied slightly outside of the $99\%$ confidence interval. As the discrepancy became more pronounced for lower lying zeros (around $10^{16}$-th zero) of the Riemann zeta function, we attribute a mismatch to the nonuniversal finite energy effects.
\end{references}
\begin{references}
\bibitem{Rem-unity} More generally, it holds: $\langle s_\ell^{\rm loc}/s_0^{\rm loc}\rangle = 1$ for all $\ell \ge 1$.
\end{references}
\end{document}